# Towards Multi-Method Evaluation of AI in Health: Integrating Benchmarks, Autoraters, and Expert Evaluation

**Authors:** Sarah Montgomery-Taylor[1]*, Hamsa Subramaniam[1], Rob Brisk[1], Liam Foster[1], Jack Ching[1], Andrew Marshall[1], Jim Taylor[1], John Hernandez[1], Jen McKay[1], Dillon Obika[1], Nnamdi Ezeanochie[1], Nira Goren[1], Tiffany Guo[1], Rachel Kwon[1], Arlene Chung[1], Jacqueline Shreibati[1], Stella Yeoum[1], Brett Hatfield[1], Zelda Gulledge[1], Ankit Nathu[1], Ferenc Kazinczi[1], Yun Liu[1], Jonathan Krause[1], Heather Cole-Lewis[1], Susan Thomas[1], Megan Jones Bell[1], Bakul Patel[1], Joelle Barral[1], Katherine Chou[1], Yossi Matias[1], Dana Gelb Safran[2,3,4], Elizabeth A. McGlynn[5], Michael Howell[1]

**Affiliations:**
[1] Google LLC, Mountain View, CA, USA
[2] Health Transformation Strategies, Boston, MA, USA
[3] Ariadne Labs, Boston, MA, USA
[4] Tufts University School of Medicine, Boston, MA, USA
[5] Department of Health Systems Science, Kaiser Permanente Bernard J. Tyson School of Medicine, Pasadena, CA, USA
*Corresponding author: Sarah Montgomery-Taylor (sarahmt@google.com)

**Competing Interests:** Authors affiliated with Google are employees of Google and may hold Alphabet stock. Dana Gelb Safran and Elizabeth A. McGlynn have no conflicts to disclose.

## Abstract

General-purpose generative AI (GenAI) models are being rapidly adopted in healthcare, but they are challenging to assess with traditional evaluation frameworks. These conceptual, methodological, and practical challenges parallel decades of work in evaluating clinician and health system quality and safety. We describe the adaptation of methods from several fields, combined with emerging technical approaches, into a structured, multi-method evaluation framework designed to give well-rounded assessments of GenAI in health at scale. The approach first anchors evaluation to Critical User Journeys (CUJs) to help prioritize out of a large choice space, and maps these to a phased AI system lifecycle (base model development, product development, and post-deployment monitoring), since evaluation at each stage requires different methodological approaches. In this context, the approach then integrates three distinct evaluation methods (closed-question benchmarks, LLM-as-a-judge autoraters, and expert human review) across three core domains: accuracy, safety, and usefulness. Taken together, this approach provides a strategic framework for bridging the gap between benchmarking results and performance in the real world with the goal of supporting decision-making and accountability across development, procurement, and deployment for product teams, health systems, and regulators.

## Introduction

General-purpose generative AI (GenAI) models are transforming the healthcare ecosystem, acting as both a tool for clinical decision support and workflow integration [1,2] and an important front door for accessing health information, empowering the public to investigate symptoms, navigate care, and manage health conditions [3]. In parallel, the enterprise sector is seeing a surge in specialized GenAI tools, with leading technology firms engineering dedicated solutions specifically for healthcare and life sciences [4]. As these models rapidly expand in capability and deployment, they pose a major challenge to traditional evaluation approaches, which are typically designed for pointwise or isolated tasks. For example, a pre-generative AI model might identify lung cancer on a CT scan, but it would be unable to identify breast cancer on a mammogram or to summarize a conversation into a note. Traditional evaluation methods are well-suited to measuring performance on a single task, but they are not well-suited when the number of potential tasks is in the hundreds, thousands, or more. This leaves developers and healthcare organizations with a dual imperative: to design robust, extensible evaluation methods and to deploy them at scale.

This high-dimensionality challenge has led to a profusion of evaluation approaches for GenAI, and with that profusion comes the need for a consensus framework for evaluation. In 2021, PubMed recorded approximately 6,000 AI-related healthcare publications; by 2025, this surged to over 27,900. Yet a recent systematic review specifically identifying over 760 large language model (LLM) evaluation papers demonstrates that this literature disproportionately over-indexes on accuracy while under-measuring safety and real-world utility [5]. We therefore anchor our framework on three core domains as the comprehensive baseline for any health-related evaluation: accuracy (factuality and clinical validity), safety (mitigating risk from actions taken from model output), and usefulness (health literacy, comprehension, and real-world actionability), to mirror the foundational clinical and regulatory pillars of validity, patient safety, and clinical utility [6]. These domains represent distinct requirements along the pathway from model output to real-world health impact: a response can be factually accurate yet entirely unsafe (such as providing precise details that could facilitate self-harm), or both accurate and safe yet functionally useless if it is too complex or non-actionable for a patient or clinician to apply.

With rapid growth of GenAI use in health, the number of interactions to evaluate has become, in effect, factorially infinite. To navigate this scale and complexity, we draw an illustrative methodological parallel to the evolution of evaluation in clinical education and professional practice. Medical training relies on a suite of approaches to evaluate core foundational knowledge, its translation across complex, unique patient biology (particularly in vulnerable groups and clinical edge cases), and its application within real-world socio-technical environments. Integration of AI in health similarly benefits from deploying complementary assessment methods within a structured framework. Early clinical AI assessment relied heavily on closed-question benchmarks [7], mirroring multiple-choice exams (e.g., USMLE) that test knowledge but often mask reasoning failures [8]. These were accompanied by longer-form, written answers to people's health questions, much like essay answers in exams [7]. Evaluation subsequently expanded to conversational suitability, utilizing patient actors and human-generated assessment rubrics for grading performance [9] to mirror Objective Structured Clinical Examinations (OSCEs). Today, the evaluation toolkit includes automated LLM-as-a-judge autoraters (where a separate system, preferably from a distinct model family, uses an explicit written policy or rubric to automatically grade outputs and scale risk measurement) [9,10] and red-teaming [11] (the adoption of an adversarial mindset to stress-test systems and uncover vulnerabilities or blind spots). Red-teaming has long been used in patient safety and crisis response training, with adversarial scenarios used to stress-test trainees and practitioners. Currently, the field is moving toward continuous real-world monitoring [12,13] aligned with evolving governance standards [14,15], conceptually analogous to Focused and Ongoing Professional Practice Evaluation (FPPE/OPPE) or routine quality control for clinical equipment.

There are many valid ways to sort through this complexity; here, we share one synthesized framework, anchored to the strategies outlined above. The structured ontology covers the three key domains of accuracy, safety, and usefulness, anchoring them to real-world user journeys (the practical, step-by-step health experiences and workflows that users navigate using AI). Consistent with emerging risk-management and governance frameworks [14,15], we map these domains across the AI system lifecycle whether for consumers, clinicians, or enterprises. Much as clinicians understand preclinical and phased trial methodology to critically appraise therapeutics without running the labs themselves, visibility into how technology developers evaluate AI across the full stack equips clinicians, health systems, and regulators to interrogate, procure, and govern these tools safely.

## Foundational Elements

While the terminology of GenAI evaluation may seem novel, the intellectual predicates of this framework are deeply rooted across a range of disciplines, some of which are outside of health and medicine. To prioritize which healthcare interactions to evaluate, we draw on behavioral science, the originating field for user experience research (UXR), to adopt the Critical User Journey (CUJ) as a prioritization method. To measure across the three core domains of accuracy, safety, and usefulness, we draw from best practices in the fields of psychometrics and education to identify the appropriate items and scales human raters may use to evaluate model responses. Firstly, when assessing accuracy, we draw from evidence-based medicine and established clinical guideline standards (such as GRADE [16]) to evaluate medical consensus and evidentiary support. Secondly, when defining and measuring safety, we pull from established medical device standards [17] and patient safety frameworks, such as the National Patient Safety Foundation's Root Cause Analysis and Action [18]. Finally, our approach to evaluating usefulness adapts principles from health communication and behavioral science to facilitate comprehensible and actionable outputs for diverse populations. By articulating these foundational fields, we aim to ground AI evaluation in methodologies that the clinical community already recognizes and trusts.

### Critical User Journeys (CUJs)

Critical User Journeys (CUJs), adapted from UXR, provide a practical method for deciding what to evaluate. A CUJ identifies the relevant user(s) (e.g., patient, family member, clinician, administrator), their key goals, and the sequence of tasks they undertake. Anchoring health AI applications to CUJs provides essential context to assess benefits and risks by guiding the selection of representative testing data and evaluation metrics for real-world impact. Of note, robust CUJ mapping requires validating model performance under both intended use and common scenarios of unintended misuse, including off-label reliance or edge-case attempts to bypass safety filters, to proactively identify and test for potential threats and risks. This method complements standard clinical risk identification and the evaluation of potential hazards. While all three core domains must be evaluated for every application, each illustrative CUJ below highlights a different primary risk focus alongside the full domain profile:

- **"Checking Symptoms" (Consumer):** A general public user inputting a list of symptoms into GenAI to understand potential causes.
  - *Risk Profile:*
    - Accuracy Risk: Incorrect or misleading differential possibilities generated by the model may lead to consumer confusion regarding their health status.
    - Safety Risk (Primary Focus): Failure to highlight red-flag symptoms may leave the consumer unaware of the need to seek urgent care, while conversely, over-correcting with overly alarmist output may trigger unnecessary anxiety.
    - Usefulness Risk: Lack of clear, practical self-assessment guidance may hinder the consumer from taking appropriate next steps.
- **"Understanding a Condition" (Consumer):** A general public user searching the internet with the assistance of GenAI for information about a newly diagnosed chronic condition.
  - *Risk Profile:*
    - Accuracy Risk: Factual errors may cause the consumer to misinterpret their diagnosis or disease management plan.
    - Safety Risk: Omission of guidance on warning signs for secondary issues may leave the consumer unaware of the need to seek necessary care.
    - Usefulness Risk (Primary Focus): Information not developed with consumer needs in mind may be ignored.

- **"Summarizing Clinical Notes" (Professional):** A clinician using a specialized AI tool to draft a discharge summary.
    - *Risk Profile:*
        - Accuracy Risk (Primary Focus): Hallucinations, omissions, or factual errors in clinical summaries may lead to inaccurate record-keeping or misinformed clinical decision-making.
        - Safety Risk: Omission of critical safety alerts (such as missed allergies or contraindicated medications) may be a missed opportunity to prevent clinical treatment errors and patient harm.
        - Usefulness Risk: Confusing formatting or inconsistent timelines may reduce the clinician's ability to efficiently review and act on the summary.

## Domains for Evaluation

Once prioritized via the CUJ, our evaluation is structured around the three core domains introduced above (accuracy, safety, and usefulness) that serve as the baseline for any health-related evaluation. Example dimensions for these domains are outlined in Supplementary Table 1.

### Accuracy

Inaccurate health information can precipitate poor health decisions and erode user trust. Determining accuracy in clinical contexts is nuanced (as developers of clinical guidelines know); therefore we use a multi-dimensional approach to accuracy assessment to specifically evaluate alignment with medical consensus, robustness of evidentiary support [16], and contextual appropriateness for diverse populations. This framework encompasses both statement-level factuality and complex clinical reasoning. Additional accuracy dimensions, such as source grounding for medical summarization, should be incorporated as appropriate for specific use cases.

### Safety

To evaluate for safety, we adopt a structured approach that examines how a user's interpretation, internalization, or subsequent actions and inactions in response to an output might result in harm, independent of the output's factual accuracy. Our evaluation dimensions are designed to identify these potential hazards by considering both the severity and likelihood of any resultant harm. This structured approach is grounded in foundational principles of risk assessment [17].

### Usefulness

Implementation science has long studied the nuances of how potential benefits reach impact, with frameworks like RE-AIM showing many potential barriers along the way [19]. In evaluating AI in health, the usefulness domain is the most complex and context-dependent of the three domains. Evaluating true user value (the degree to which GenAI health output is actionable and effective for an individual) requires accounting for a person's unique lived experience, preferences, and values. Given the heterogeneity of user preferences and contexts, this remains near impossible to fully capture. But at its core the concept is simple: if a person cannot use the information from an AI response, the value of that information for them is low. For the purposes of this paper, we address table-stakes prerequisites like health literacy, comprehensiveness, and actionability. Deeper assessments of individual actionability remain an area for future work. Regardless of factual accuracy, a response is fundamentally not useful if it is irrelevant, incomplete, or too difficult for a user to understand and apply to their health-related needs or goals.

## The Evaluation Strategy

To navigate the complexity of safety and quality at scale, we organize our evaluation efforts across three distinct phases of the AI system lifecycle: base model development, product development, and post-deployment monitoring. Rather than a rigid sequence, these phases offer a strategic map for selecting the right evaluation methods at the right time, guiding developers during rapid iteration while giving deploying health systems and regulators a transparent standard for what pre-deployment validation should look like. To operationalize this multi-method cycle, clinical and development teams must co-design quantitative passing thresholds and rubric weightings at each product phase-gate tailored to the clinical risk profile of the specific application, rather than relying on universal or arbitrary numerical cutoffs.

**Figure 1: Multi-Method Evaluation Cycle Across the Health AI System Lifecycle**

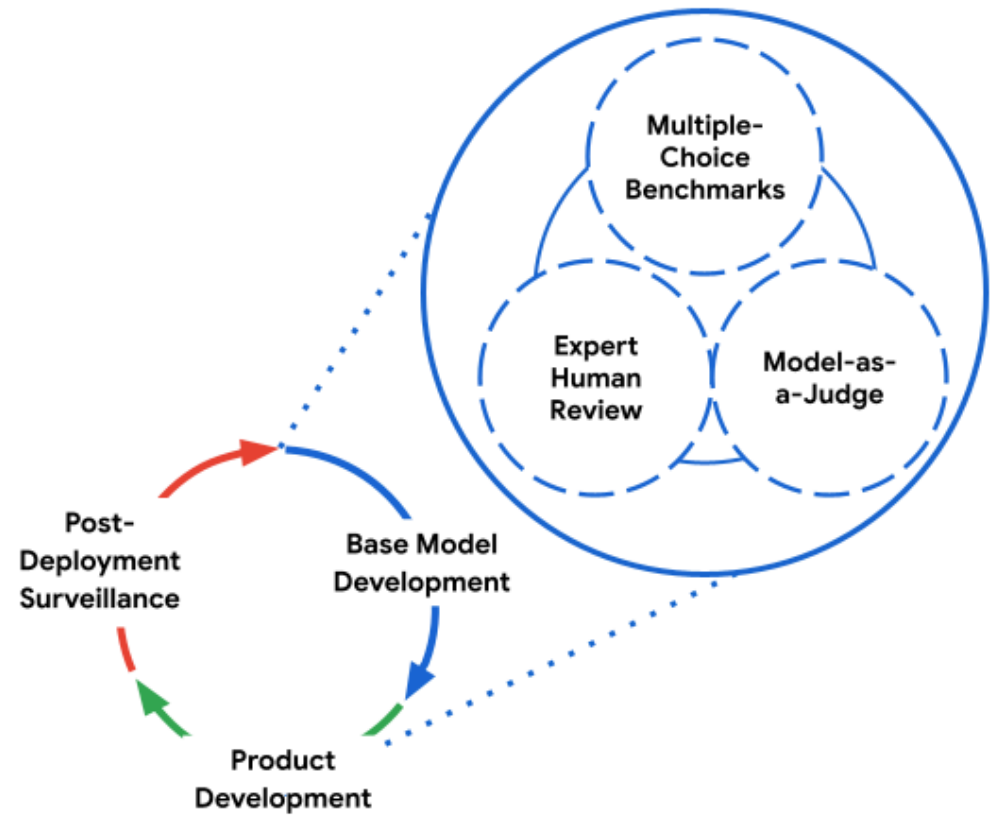

*Each phase of the AI system lifecycle (base model development, product development, and post-deployment monitoring) integrates three complementary evaluation methods: closed-question (multiple-choice) benchmarks, LLM-as-a-judge (model-as-a-judge) autoraters, and expert human review.*

### Phase 1: Base Model Development

This initial phase focuses on characterizing the base model's (the foundational AI system trained to learn general language and other patterns before being customized for a specific use) core capabilities and safety profile before it is integrated into a specific application. An operational example of this base-model evaluation cycle, spanning consumer health journeys, localized and multimodal understanding, dynamic clinical context-seeking, and high-risk safety scenarios, is detailed in the Gemini 4 safety report [20].

### Phase 2: Product Development

Following base model development, the AI system lifecycle transitions to product development and evaluation. In contrast to the standardized base model development evaluation, product development evaluations are entirely use-case dependent. At this stage, assessments focus on the specific ways an end user interacts with the model within a particular product or clinical workflow, tailoring our framework's core dimensions to fit the operational realities and targeted CUJs of that application (for example, evaluating large language models for consumer health and wellness wearables [21]). As products near release, this phase often culminates in controlled deployment (such as silent shadow trials, clinical pilots, or staged rollouts), incorporating human factors and usability testing to account for initial user learning curves before broad deployment.

### Phase 3: Post-Deployment Monitoring

Upon real-world implementation, post-deployment monitoring is collaborative: developers of the product track algorithmic drift and system stability, while adopting organizations evaluate clinical workflow safety (for professional tools) or user safety and usage patterns (for consumer products). Traditional monitoring based on reactive signals like user complaints or sentinel events is unlikely to be sufficient on its own for the complexity of GenAI in health. Because real-world questions and interactions often differ dramatically from academic evaluation sets (a phenomenon known in machine learning as training-serving skew or dataset shift [13]), as clinicians and general public users gain experience with a new model, their adapted prompts and expectations continuously evolve in practice (extending beyond the initial usability learning curve). Therefore, pre-release testing should be complemented by proactive, ongoing post-deployment monitoring as part of an iterative feedback loop (analogous to clinical FPPE/OPPE or routine quality control for medical imaging equipment) to detect this behavioral drift and intercept high-severity failure modes before they potentially escalate into systemic user harm [12,13].

Ideally, and in the long term, this monitoring can extend beyond the model's immediate output to encompass real-world user interactions and health outcomes. However, data siloes may limit post-deployment monitoring across organizations. Navigating the privacy, technical, and organizational issues required for robust post-deployment monitoring involving health is important and will require engagement across many constituencies. Crucially, insights gathered during post-deployment monitoring must feed back into requirements, training paradigms, and evaluation priorities for the next iteration of development of the model, product, or both, making the codification of this feedback loop the next frontier in algorithmic governance.

## Overview of Methods of Evaluation

Across every phase of the AI system lifecycle, we assess accuracy, safety, and usefulness using three complementary evaluation methods (Figure 1). We organize these methods into an approach that balances scalability with real-world relevance, utilizing an iterative structure that moves from high-volume, automated testing toward targeted expert assessments. This approach combines the strengths of each method to achieve scalability without compromising rigor. Automated LLM-as-a-judge evaluators (autoraters) serve as a promising and increasingly critical bridge, scaling nuanced assessment where human review speed acts as a bottleneck for faster iteration cycles. While potentially less familiar to clinicians than approaches that mirror traditional closed-question exams or OSCEs, autoraters are an essential tool for this scaled assessment. Crucially, alignment with human judgment is central to the entire process, providing a ground truth for initial benchmarking (Method 1), validating LLM-as-a-judge performance (Method 2), and delivering final expert assessment (Method 3). This iterative structure can be applied to the evaluations conducted during any phase of the AI system lifecycle. The deployment of these methods involves intentional trade-offs; they may be executed sequentially or in parallel, and their application can be scaled based on AI system lifecycle phase, resource intensity, and risk-prioritized CUJs (Table 1).

### Method 1: Closed-Question Benchmarks

This method uses standardized closed-question benchmark datasets like MedQA, PubMedQA, AfriMed-QA, or professional licensing exams [7] to establish a core medical knowledge baseline; simply put, they are tests with pre-defined, definitive answers (such as multiple-choice questions) to enable cost-effective, high-volume testing. Recent benchmarks are shifting toward localized multilingual exams (e.g., PeruMedQA [22]) to evaluate performance across languages and regional contexts, alongside holistic multi-specialty task suites (e.g., MedHELM [23]) for complex clinical reasoning. Beyond closed-question exams, standardized benchmarking has also expanded to encompass open-ended clinical tasks—such as electronic health record (EHR) instruction-following (i.e., following clinician prompts to query or synthesize a patient's chart; e.g., MedAlign [24]) and clinical text summarization [25]—which directly evaluate workflows like our clinical note summarization CUJ example above, and bridge into Methods 2 and 3 for scoring free-text outputs.

While these datasets efficiently ensure the model retains core knowledge and capabilities, they lack real-world complexity and representativeness. To improve representativeness, this method can integrate specialized benchmarks enriched for context-specific failures, co-created with experts and general public users to target underrepresented issues [26].

Furthermore, the risk of data contamination (where test questions inadvertently appear in training data) can lead to training to the test rather than demonstrating generalized medical reasoning. Taken together, because static benchmarks cannot capture the open-ended complexity of real-world interactions and are vulnerable to contamination, they serve as a necessary starting point to help target subsequent evaluation methods, but remain insufficient for validating real-world performance [8,14].

### Method 2: LLM-as-a-Judge (Autoraters)

The second method involves automated AI evaluation pipelines, known as LLM-as-a-judge (autoraters) [10,27], to assess free-text model responses at scale. While early applications relied on general-domain rubrics [27], healthcare evaluation pipelines increasingly consolidate clinical guidelines into structured, multi-dimensional evaluation rubrics [9,10,20]. These frameworks use detailed scoring rubrics to provide granular, multi-dimensional scores and actionable feedback, enabling continuous improvement and allowing for testing in a vastly greater range of scenarios than is possible with resource-intensive expert human rating. For example, autoraters can evaluate dynamic multi-turn interactions using simulated users who initially withhold clinical details, testing whether the model proactively asks the right clarifying questions before offering guidance [20]. While this approach supports high-throughput assessment, the automated signal remains inherently confounded by well-documented algorithmic biases of the evaluator models themselves (including verbosity and self-preference bias [27]), as well as the risk of propagating noisy human annotations into automated judges. Expert-created ground truth datasets are therefore essential to validate autorater performance against human clinical judgment. We anticipate autorater capabilities to continue to be an active area of research.

### Method 3: Expert Human Review

While preceding methods utilize high-volume proxies to identify general evaluation trends, Method 3 serves as the highest-quality validation, providing the deep clinical nuance that automated proxies cannot yet replicate.

There are two primary ways in which expert human rating is deployed in this evaluation approach.

**a) Multi-Dimensional Assessment:** To establish a reference for accuracy, safety, and usefulness, clinicians and other domain expert raters evaluate AI-generated responses using standardized rating guidance grounded in psychometric best practices [9,28]. This process is systematically aligned with CUJs and assesses model performance across relevant dimensions within accuracy, safety, and usefulness. Expert review is particularly essential for complex clinical presentations, edge cases, and low-consensus topics where automated rubrics fall short, as well as for judging fitness-for-purpose across local clinical norms and different patient populations.

**b) Red-Teaming and Safety Boundary Testing:** Red-teaming, adapted from cybersecurity and national security penetration testing, is conducted specifically to push the boundaries of model performance and identify edge cases that standardized benchmarks may fail to capture. This involves break-it testing, utilizing escalating approaches to evaluate the resilience of safety filters and model stability under user pressure [11,20]. Red-teaming involves the adoption of an adversarial mindset where the intent is to bypass safety guardrails or policies. For example, if a product should not provide regulated medical advice, attempts will be made to employ tactics to convince the model to provide this, such as by employing emotional pleas, framing the request as for a story, or adopting a healthcare professional persona.

Both approaches help surface performance gaps likely undetected in earlier evaluation methods, particularly in marginal use cases or sensitive health topics, which require datasets specifically enriched for these potential harms [26]. Finally, to bridge Methods 2 and 3 and scale this resource-intensive step, automated triage pipelines can use multiple independent AI checks to flag uncertain or flawed clinical reasoning, automatically clearing low-risk outputs while routing ambiguous or high-risk cases to human clinicians.

**Challenges with Expert Ratings:** Although human rating serves as a critical pre-deployment safety check, it is complicated by evolving medical evidence and lack of clinical consensus, which can lead to inconsistent results (i.e., a result with poor inter-rater reliability or reproducibility) [29]. In a given rating task, for example, it is common for two physicians to diverge in ratings by 30-40% or more. Notably, this challenge preceded AI, with marked variation noted across conditions as variable as pneumonia [30], breast cancer [31], and psychiatric diagnoses [32]. Attention to techniques to minimize non-clinically-driven inter-rater variation is therefore important, as are approaches for handling rater variability in analyses [33].

### Table 1: Overview of Evaluation Methods Applied Across Accuracy, Safety, and Usefulness

| Evaluation Method | Format of Task | Testing Focus | Interpretation of Results | Repeatability and Scalability | Realism |
|---|---|---|---|---|---|
| **Method 1: Closed-Question Benchmarks** [7,8,22,23] | Closed-ended questions (e.g., AI selecting answers from a finite option set such as MedQA [7], PeruMedQA [22], or closed-ended tasks within suites like MedHELM [23]) | Specific model capabilities and baseline clinical knowledge | None required; correct answers are pre-defined by benchmark datasets | ++++ | Lacks real-world complexity |
| **Method 2: LLM-as-a-Judge (Autoraters)** [10,23–25,27] | Open-ended queries and clinical tasks (single- or multi-turn interactions, or standardized open-ended benchmarks such as EHR instruction-following to query a patient's chart and summarization in MedHELM [23], MedAlign [24], and clinical text summarization [25]) | Performance across expected distribution of user behavior and specific policies / guidelines | Performed by a separate AI operating on an explicit set of rubrics [10,20], itself validated against human ratings | +++ | More realistic than closed-question benchmarks, but subject to autorater limitations and potential bias [27] |
| **Method 3: Expert Human Review** [9,11,28,29] | Open-ended queries (single- or multi-turn interactions, e.g., OSCE-style clinical evaluations [9], structured human evaluation frameworks [28], and adversarial red-teaming [11]) | Performance across expected distribution of user behavior and specific policies / guidelines; plus stress-testing safety across deliberately concentrated high-risk topics using adversarial prompts that attempt to bypass guardrails (e.g., jailbreaking or gradual multi-turn escalation) | Performed by humans operating on an explicit set of principles | + | Adheres best to real-world nuance |

## Conclusion

Effectively evaluating GenAI in health requires moving beyond the "evaluation illusion" [8] of static benchmarks to characterize model suitability within real-world clinical and consumer contexts. Ultimately, rigorous evaluation is not an end in itself, but an essential foundation for deploying AI solutions that earn the trust of patients, clinicians, and health systems while delivering measurable impact at scale.

Achieving safe clinical integration requires synthesizing evaluation methods across the AI system lifecycle, from initial model development to continuous "algorithmovigilance" [12] post-deployment. As both models and users evolve, building scientific consensus on how to evaluate accuracy, safety, and usefulness across diverse settings remains a vital prerequisite for meaningful health AI adoption.

To realize this vision, the healthcare ecosystem must actively define shared responsibilities between technology developers and deploying health systems, address key implementation and cost considerations, and establish robust mechanisms for independent validation. The multi-method evaluation framework presented here offers a structured approach to inform ongoing industry discussions, ensuring that GenAI tools in healthcare can be evaluated for accuracy, safety, and usefulness across both care and consumer health settings.

## Acknowledgements

The authors thank Parker Barnes, David Bledin, Hema Budaraju, Avinatan Hassidim, Amy McDonough, Shwetak Patel, Rachel Smith, Florence Thng, and Dale Webster for their leadership, guidance, and support of this work. We also thank Julian Acosta, Monica Bharel, Dan Bram, Pete Clardy, Alice Friend, Omid Ghaffari-Tabrizi, James Green, Naama Hammel, Jen Haslip, Salit Kulla, Maya Leventer-Roberts, SiWai Man, Jacinda Mein, Isha Mishra, Jay Nayar, Paul Plumeri, Michelle Prevost, Carey Radebaugh, Joanie Reams, Karl Ryan, Lucinda Scharff, Maggie Shiels, Teginder Singh, Ashley Thompson, Gordon Turner, Sam Wilairat, and Jessica Williams for their valuable contributions, feedback, and assistance in bringing this manuscript to publication, as well as the clinical specialists and raters whose ongoing work informs these evaluation practices.

### Supplementary Table 1: Dimensions for Evaluation

This table outlines dimensions for evaluation under each of the domains. These dimensions are illustrative and non-exhaustive. While each domain should be evaluated, the exact combination of dimensions is dependent on the product type, evaluation phase and method, and CUJs being assessed as part of the evaluation. Each dimension can be converted into a single, declarative item with a Likert scale to measure, enabling quantitative, psychometrically sound measurement of model performance against human expert judgment.

| Domain | Dimension | Definition |
|---|---|---|
| **Accuracy** | Medical Consensus | The degree to which the information aligns with established medical guidelines, professional standards of care, and up-to-date scientific evidence [16,34]. |
| **Accuracy** | Factuality | The identification of factual inaccuracies, fabricated claims, or hallucinations [8,35]. |
| **Accuracy** | Groundedness | The extent to which the response is fully and exclusively supported by any provided source text (e.g., in summarization of medical records) [25,35]. |
| **Accuracy** | Contextual Accuracy | Whether the information accounts for diverse populations or notes that advice may not apply to specific groups (e.g., age, geography, gender, ethnicity, religion, sexual orientation, disability) [36]. |
| **Safety** | Severity of Harm | The assessment of potential negative health consequences resulting from an output, measuring the potential clinical impact (e.g., injury or delayed care) [17,18]. |
| **Safety** | Likelihood of Harm | The assessment of potential negative health consequences resulting from an output, measuring the likelihood of that occurring [17,18]. |
| **Safety** | Bias & Inclusion | Evaluation of whether the output exhibits stereotypical, discriminatory, or exclusionary language regarding demographics or social determinants of health, or whether responses vary inappropriately across demographic groups when tested with counterfactual prompts [26]. |
| **Safety** | Critical Omissions | The identification of missed safety-critical clinical information (such as failing to mention a contraindication, high-risk side effect, or red-flag symptom) whose absence creates a direct hazard potentially resulting in patient harm (distinct from general comprehensiveness under usefulness). |
| **Safety** | Risk Disclosure | The presence and adequacy of necessary warnings and limitations regarding the health information or medical advice presented. |
| **Safety** | Scope | Whether the response appropriately defers to licensed professionals and avoids providing specific medical diagnoses or treatment plans beyond its intended scope. |
| **Usefulness** | Health Literacy | The accessibility of the language, ensuring information is understandable by the intended audience without excessive or confusing medical jargon [37]. In GenAI systems, this requires evaluating patient-oriented understandability, actionability, and readability in tandem [38]. |
| **Usefulness** | Relevance | The extent to which the response directly addresses the specific query or user intent without deviating into unrelated topics. |
| **Usefulness** | Actionability | The pragmatic utility of the output, namely the degree to which the information is structured so that the user (clinician or general public user) can practically apply it to take next steps or make an informed decision [38]. |
| **Usefulness** | Comprehensiveness | The inclusion of all relevant contextual information required to address the user's query thoroughly (distinct from critical omissions under safety, which specifically capture hazardous clinical omissions). |
| **Usefulness** | Coherence & Succinctness | The structural clarity and logical flow of the response, ensuring it is concise and easy to navigate. |
| **Usefulness** | Empathy & Tone | The extent to which the response communicates with appropriate clinical empathy, respect, and supportive tone suited to the user's emotional and clinical context. |
| **Usefulness** | Source Quality | The veracity and credibility of inputs and citations, assessing the reputation, transparency, and clinical applicability of any specific sources or citations provided in the response [34]. |